\documentclass[apjl]{emulateapj}

\usepackage{graphicx}

\usepackage{lineno}

\usepackage{subfigure}
\usepackage{color}

\usepackage{color}
\usepackage{amssymb}
\usepackage{amsmath}
\usepackage{enumerate}

          \font\sixrm=cmr6       
                    
\def\edithere#1{\textcolor{black}{#1}}  

\def\fermi{{\it Fermi}}
\newcommand{\Mesz}{{M\'esz\'aros}}
\def\mathnew{\mathsurround=0pt}
\def\simov#1#2{\lower .5pt\vbox{\baselineskip0pt \lineskip-.5pt
      \ialign{$\mathnew#1\hfil##\hfil$\crcr#2\crcr\sim\crcr}}}

\def\beq{\begin{equation}}
\def\enq{\end{equation}}
\def\bea{\begin{eqnarray}}
\def\ena{\end{eqnarray}}

\def\L54{L_{54}}
\def\E55{E_{55}}
\def\et3{\eta_3}
\def\th1{\theta_{-1}}
\def\r07{r_{0,7}}
\def\x05{x_{0.5}}
\def\et600{\eta_{600}}
\def\et3{\eta_3}

\def\gamth{\gamma_{\rm th}}
\def\gammaMin{\gamma_{\rm min}}

\def\vFv{\nu F_{\nu}}

\def\mathnew{\mathsurround=0pt}
\def\simov#1#2{\lower .5pt\vbox{\baselineskip0pt \lineskip-.5pt
       \ialign{$\mathnew#1\hfil##\hfil$\crcr#2\crcr\sim\crcr}}}

\def\beq{\begin{equation}}
\def\enq{\end{equation}}
\def\bea{\begin{eqnarray}}
\def\ena{\end{eqnarray}}

\def\L54{L_{54}}
\def\E55{E_{55}}
\def\et3{\eta_3}
\def\th1{\theta_{-1}}
\def\r07{r_{0,7}}
\def\x05{x_{0.5}}
\def\et600{\eta_{600}}
\def\et3{\eta_3}

\defcitealias{Veres+12magnetic}{VM12}

\def\dover#1#2{\hbox{${{\displaystyle#1 \vphantom{(} }\over{
\displaystyle #2 \vphantom{(} }}$}}
{\catcode`\@=11                                                  
\gdef\SchlangeUnter#1#2{\lower2pt\vbox{\baselineskip 0pt\lineskip0pt    
\ialign{$\m@th#1\hfil##\hfil$\crcr#2\crcr\sim\crcr}}}}

\def\gamth{\gamma_{\hbox{\sixrm th}}}
\def\gammaMin{\gamma_{\hbox{\sixrm min}}}

\def\vFv{\nu F_{\nu}}

\begin{document} 
\title{An Observed Correlation Between Thermal and Non-Thermal Emission in Gamma-Ray 
Bursts}

\author{J. Michael Burgess\altaffilmark{1,2},
Robert D. Preece\altaffilmark{1,2},
Felix Ryde\altaffilmark{3,4},
Peter Veres\altaffilmark{5}, 
Peter \Mesz\altaffilmark{5}, 
Valerie Connaughton\altaffilmark{2}, 
Michael Briggs\altaffilmark{2}, 
Asaf Pe'er\altaffilmark{6}, 
Shabnam~Iyyani\altaffilmark{4,3,7},
Adam Goldstein\altaffilmark{8},
Magnus~Axelsson\altaffilmark{9,3,4},
Matthew~G.~Baring\altaffilmark{10}, 
P. N. Bhat\altaffilmark{2}, 
David Byrne\altaffilmark{11}, 
Gerard Fitzpatrick\altaffilmark{11}, 
Suzanne Foley\altaffilmark{11,12},
Daniel~Kocevski\altaffilmark{13}, 
Nicola~Omodei\altaffilmark{13},
William S. Paciesas\altaffilmark{14}, 
Veronique Pelassa\altaffilmark{2}, 
Chryssa Kouveliotou\altaffilmark{7},
Shaolin Xiong\altaffilmark{2}, 
Hoi-Fung Yu\altaffilmark{12}, 
Binbin Zhang\altaffilmark{2}, 
Sylvia Zhu\altaffilmark{15}
}

\altaffiltext{1}{Department of Space Science, University of Alabama in
  Huntsville, Huntsville, AL 35899, USA:\\jmichaelburgess@gmail.com,
  rob.preece@nasa.gov}

\altaffiltext{2}{Center for Space Plasma and Aeronomic Research
  (CSPAR), University of Alabama in Huntsville, Huntsville, AL 35899,
  USA}

\altaffiltext{3}{Department of Physics, Royal Institute of Technology
  (KTH), AlbaNova, SE-106 91 Stockholm, Sweden:\\felix@particle.kth.se}

\altaffiltext{4}{The Oskar Klein Centre for Cosmoparticle Physics,
  AlbaNova, SE-106 91 Stockholm, Sweden} 

\altaffiltext{5}{Department of Astronomy and Astrophysics,
  Pennsylvania State University, University Park, PA 16802,
  USA:\\veres@gwu.edu, npp@astro.psu.edu}

\altaffiltext{6}{Physics Department, University College Cork, Cork, Ireland}

\altaffiltext{7}{Department of Physics, Stockholm University,
  AlbaNova, SE-106 91 Stockholm, Sweden}

\altaffiltext{8}{Space Science Office, VP62, NASA/Marshall Space
  Flight Center, Huntsville, AL 35812, USA}

\altaffiltext{9}{Department of Astronomy, Stockholm University, SE-106
  91 Stockholm, Sweden}

\altaffiltext{10}{Rice University, Department of Physics and Astronomy,
MS-108, P. O. Box 1892, Houston, TX 77251, USA}

\altaffiltext{11}{University College Dublin, Belfield, Dublin 4, Ireland}

\altaffiltext{12}{Max-Planck-Institut f$\rm \ddot{u}$r
  extraterrestrische Physik (Giessenbachstrasse 1, 85748 Garching,
  Germany)}

\altaffiltext{13}{W. W. Hansen Experimental Physics Laboratory, Kavli
  Institute for Particle Astrophysics and Cosmology, Department of
  Physics and SLAC National Accelerator Laboratory, Stanford
  University, Stanford, CA 94305, USA}

\altaffiltext{14}{Universities Space Research Association, Huntsville,
  AL 35805, USA}

\altaffiltext{15}{Department of Physics and Department of Astronomy,
  University of Maryland, College Park, MD 20742, USA}





Double-space the manuscript.
\begin{abstract}
  

  Recent observations by the $\fermi$ Gamma-ray Space Telescope have
  confirmed the existence of thermal and non-thermal components in the
  prompt photon spectra of some Gamma-ray bursts (GRBs). Through an
  analysis of six bright Fermi GRBs, we have discovered a correlation
  between the observed photospheric and non-thermal $\gamma$-ray
  emission components of several GRBs using a physical model that has
  previously been shown to be a good fit to the Fermi data. From the
  spectral parameters of these fits we find that the characteristic
  energies, $E_{\rm p}$ and $kT$, of these two components are
  correlated via the relation $E_{\rm p} \propto T^{\alpha}$ which
  varies from GRB to GRB. We present an interpretation in which the
  value of index $\alpha$ indicates whether the jet is dominated by
  kinetic or magnetic energy. To date, this jet composition parameter
  has been assumed in the modeling of GRB outflows rather than derived
  from the data.

\end{abstract}


\section{Introduction}

Gamma-ray Bursts (GRBs) are believed to arise from the deaths of
massive stars or the coalescence of two compact stellar objects such
as neutron stars or black holes. The resulting explosion gives rise to
an expanding fireball with a jet pointed at the observer but hidden
from the observer until the density of radiation and particles in this
highly relativistic outflow is low enough for radiation to escape, a
region called the photosphere \citep[for a review
see][]{meszaros:2006}. While the emission from this fireball is
expected to be thermal \citep{Goodman:1986,Paczynski:1986},
observations over the past three decades suggest the prompt emission
to be highly non-thermal
\citep{Mazets:1981,Fenimore:1982,matz,Kaneko:2006,Goldstein:2012},
with only a few exceptions \citep{ryde:2010,ghirlanda:2013}. The
conversion of the fireball energy into non-thermal $\gamma$-ray
radiation involves the acceleration of electrons in the outflow and
their subsequent cooling via an emission process such as synchrotron
radiation \citep{sari:1998,tavani:1996}. Insight into these energy
radiation emission processes in GRBs is obtained by comparing the
observed $\gamma$-ray photon spectra directly to different radiation
models. The $\fermi$ Gamma-ray Space Telescope offers a broad energy
range for these comparisons. Recent observations
\citep{Guiriec:2010,Zhang:2011,Axelsson:2012,guiriec:2013,Iyyani:2013,Preece:2013,Burgess:2013}
show that at least two mechanisms can be present: a non-thermal
component that is consistent with synchrotron emission from
accelerated electrons in the jet and a typically smaller blackbody
contribution from the photosphere. This photospheric emission is
released when the fireball becomes optically thin so that an observer
may see a mixture of thermal and non-thermal emission with different
temporal characteristics that, when viewed together, can probe the
development and structure of the fireball jet. This simple
photospheric model has been used to quantitatively interpret
several observed correlations such as the Amati correlation
\citep[e.g.,][]{2007ApJ...666.1012T, 2011ApJ...732...34L,
  2012ApJ...755L...6F}

We are thus motivated to investigate correlations among spectral
parameters derived by fitting the non-thermal component with a
synchrotron photon model and the thermal component with a blackbody,
an approach developed in previous investigations
\citep{Burgess:2012,Burgess:2013}. The synchrotron model consists of an
accelerated electron distribution, containing a relativistic
Maxwellian and a high-energy power law tail that is convolved with the
standard synchrotron kernel
\citep{Burgess:2013,Burgess:2012,rybicki:1979}. We find that the
characteristic energies ($E_{\rm p}$ for synchrotron and $kT$ for the
blackbody) of the synchrotron and blackbody components are highly
correlated across all the GRBs in our sample. We show that this
correlation can be used to address the key question of how the energy
of the outflow is distributed, i.e., whether the energy is in a
magnetic field or is imparted as kinetic energy to baryons in the jet,
and how this energy distribution evolves with
time. 

\section{Observations}
The $\fermi$ Gamma-ray Burst Monitor (GBM) \citep{meegan:2009} has
detected more than 1200 GRBs since the start of operations on 2008,
July 14. A smaller number have been seen by the $\fermi$ Large Area
Telescope (LAT) \citep{atwood:2009} at energies greater than 100 MeV,
but these are particularly interesting because they are among the
brightest GRBs and offer the greatest opportunity for spectral
analysis across a broad energy range. GRBs can last from a few
milliseconds to hundreds of seconds or longer and have a variety of
temporal profiles, from single spikes to multi-episodic overlapping
pulses. Single-pulse GRBs exhibit the simplest spectral evolution,
providing the ``cleanest'' signal for fitting physical models to the
data \citep{Burgess:2013,Burgess:2012,Ryde:2009}.

\begin{figure}[h]
  \centering
  \subfigure[]{\includegraphics[scale=.7]{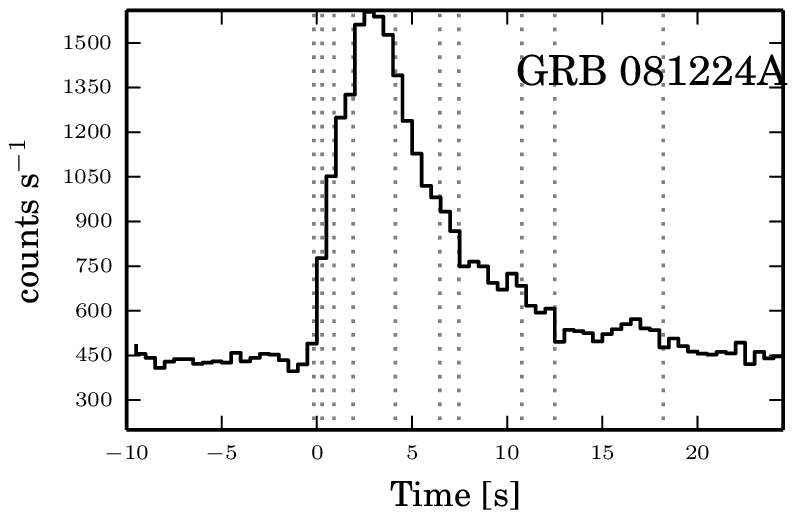}}\subfigure[]
{\includegraphics[scale=.7]{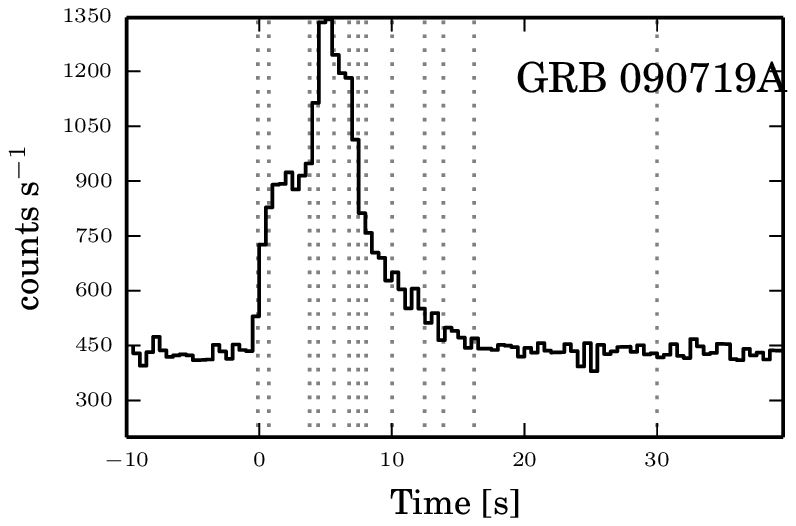}}\subfigure[]{\includegraphics[scale=.7]
{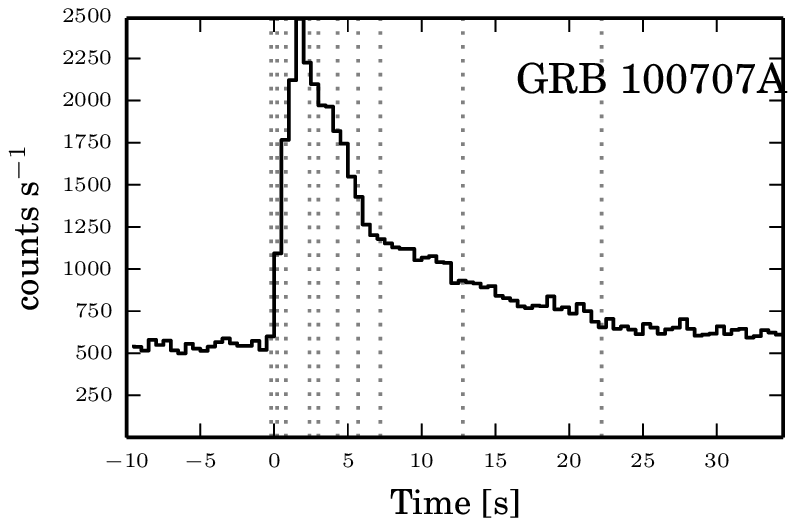}}
\subfigure[]{\includegraphics[scale=.7]{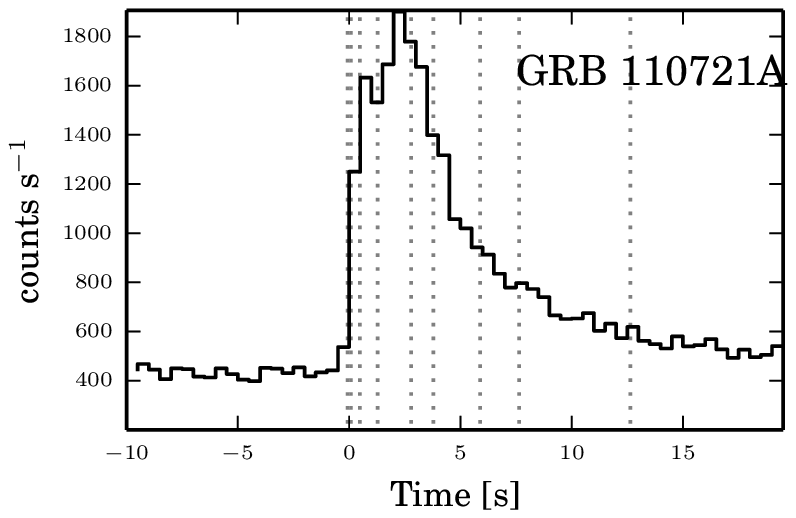}}\subfigure[]
{\includegraphics[scale=.7]{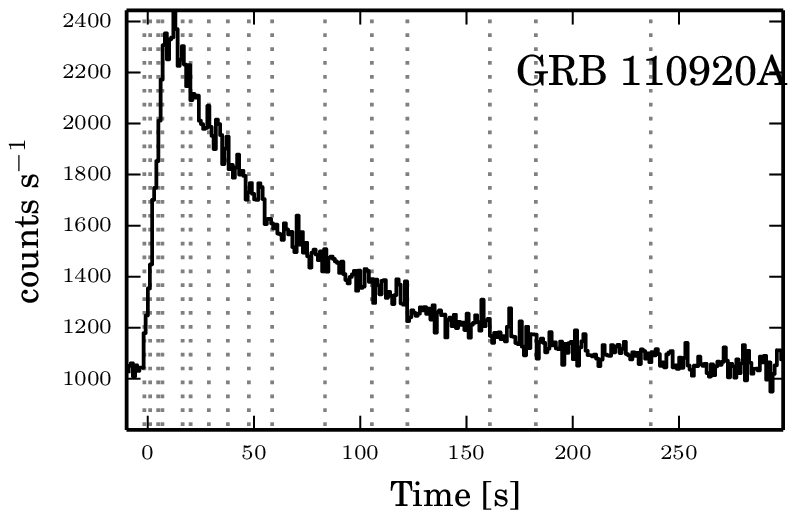}}\subfigure[]{\includegraphics[scale=.7]
{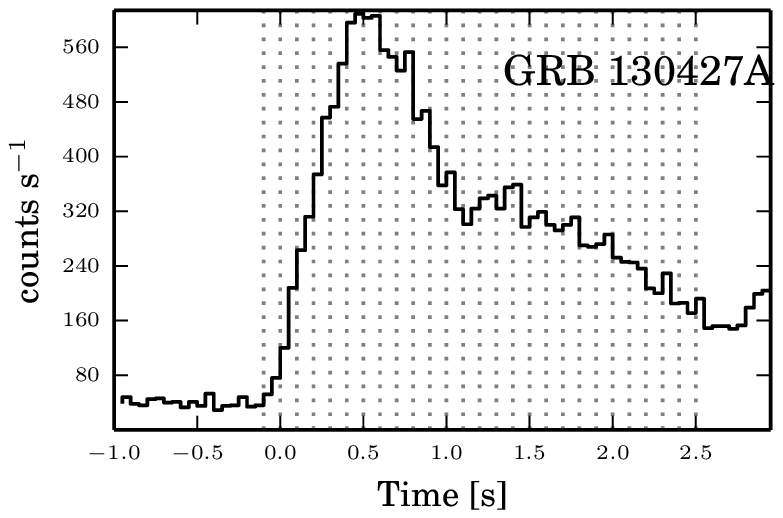}}
\caption{Each of the lightcurves in (a)-(f) shows the count rates
  detected by $\fermi$ GBM for a GRB in our sample. The
  panels show the Sodium Iodide detector count rates between 8 and 300
  keV. For each GRB in the
  sample, the time binning was selected by a Bayesian blocks algorithm
  \citep{Scargle:2013,Burgess:2013} which operates by searching for significant
  changes in the count intensity. The bin selections are indicated by
  the vertical dotted lines.}
  \label{fig:lcs}
\end{figure}

In this work, we analyze six bright, single-pulse GRBs detected by
$\fermi$ (see Table \ref{tab:tab1} and Figure \ref{fig:lcs}) and find
correlations between the $E_{\rm p}$ and $kT$ values within each of
these GRBs. The GRBs in our sample are GRB 081224A
\citep{wilson:2008}, GRB 090719A \citep{horst:2009}, GRB 100707A
\citep{wilson:2010}, GRB 110721A \citep{tierney:2011}, GRB 110920A,
GRB 130427A \citep{andreas:2013}. The time histories of these GRBs are
shown in Figure \ref{fig:lcs}, with vertical dotted lines indicating
the time binning used for the analysis of the spectral evolution of
each spectral component. In a previous analysis \citep{Burgess:2013},
the viability of fitting physical models to the $\fermi$ GRB data was
demonstrated for several GRBs and the spectral evolution of these
models over the burst durations was investigated. The synchrotron
model of \citet{Burgess:2013}, was constructed by convolving a
shock-accelerated electron distribution of the form
\begin{equation}
  n_{\rm e}(\gamma )\; =\; n_{0} \biggl\lbrack\;
  \Bigl( \dover{\gamma}{\gamth} \Bigr)^2\,
  e^{-\gamma/\gamth } + \epsilon \,
  \Bigl( \dover{\gamma}{\gamth} \Bigr)^{-\delta}\,
  \Theta \Bigl( \dover{\gamma}{\gammaMin} \Bigr)\, \biggr\rbrack\, 
  \label{eq:elec_dist}
\end{equation}
with the standard synchrotron kernel \citep{rybicki:1979}.  Here,
$n_0$ normalizes the distribution to total number or energy, $\gamma$
is the electron Lorentz factor in the fluid frame, $\gamth$ is the
thermal electron Lorentz factor, $\gammaMin$ is the minimum electron
Lorentz factor of the power-law tail, $\epsilon$ is the normalization
of the power-law, and $\delta$ is the electron spectral index. The
function $\Theta(x)$ is a step function where $\Theta(x)=0$ for $x<1$
and $\Theta(x)=1$ for $x>1$. \edithere{After convolution with the
  synchrotron kernel, the final fit parameters are the overall
  normalization of the spectrum, the $\vFv$ peak of the spectrum
  ($E_{\rm p}$), and the electron spectral index, $\delta$.} These
fits were found to be as good as those made with the empirical Band
function \citep{Band:1993} that is the common choice for GRB
spectroscopy. However, the Band function, being empirical, makes it
difficult to deduce a more physical understanding. The fits with
synchrotron model provide a direct association of the observed
spectrum with a physical emission mechanism and therefore the fit
parameters can be used to study properties of the GRB jet without
ambiguity.

All of these GRBs were shown to be consistent with a physical model
containing both a synchrotron and a blackbody component. For five of
those GRBs we investigate herein correlations between the previously
derived $E_{\rm p}$ and $kT$ values, and we add to our sample the
first pulse of the ultra-bright burst, GRB 130427A, for which a
similar analysis has been performed \citep{Preece:2013}. GRB 130427A
is the brightest GRB detected by $\fermi$ to date. Although its
temporal structure is complex \citep{zhu:2013}, it begins with a
bright single pulse that is ideal for our physical modeling, which was
used to show that internal shocks cannot explain the observed emission
\citep{Preece:2013}. GRB 081224A, GRB 110721A, and GRB 130427A were
analyzed with GBM and LAT data; the rest of the sample were analyzed
with GBM data alone. While this sample is limited by the number of
bright, single-pulsed GRBs in the $\fermi$ data set, this requirement
allows reliable interpretation of the fits without confusion from
overlapping pulses with different underlying spectra, which is
essential to measuring the evolution of the thermal and non-thermal
components throughout the duration of the GRB.


\section{A Correlation Between Spectral Components}
\begin{figure}[h]
  \centering
\includegraphics{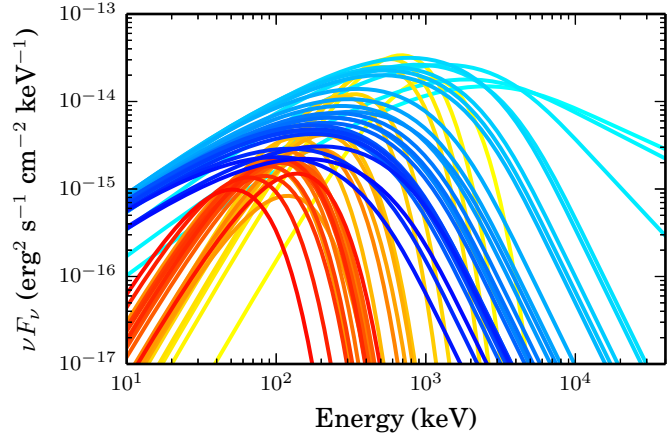}  
\caption{The $\nu F_{\nu}$ time-resolved spectrum of GRB 130427A
  \citep{Preece:2013}. The evolution of the synchrotron component
  evolves from cyan to blue while the blackbody component evolution is
  shown from yellow to red with the time bins corresponding to Figure
   \ref{fig:lcs} (f). The correlation of $E_{\rm p}$ and $kT$ is not obvious from
  the spectrum alone. Clearly the fluxes of the two components are not
  correlated.}
\label{fig:spec}
\end{figure}

\begin{figure}[h]
  \centering
  \subfigure[]{\includegraphics[scale=.7]{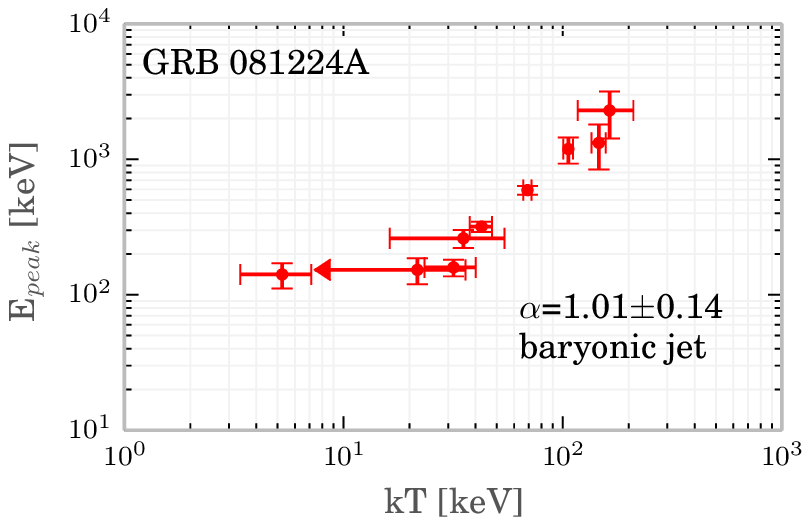}}\subfigure[]
{\includegraphics[scale=.7]{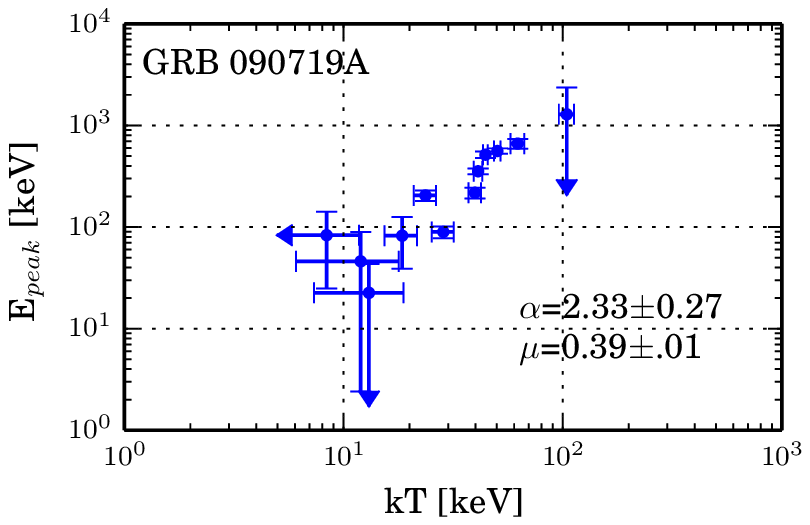}}\subfigure[]
{\includegraphics[scale=.7]{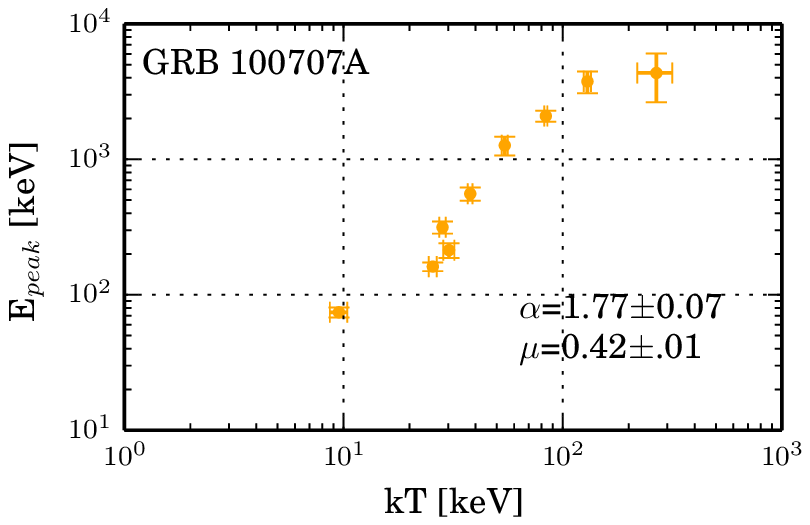}}
\subfigure[]{\includegraphics[scale=.7]{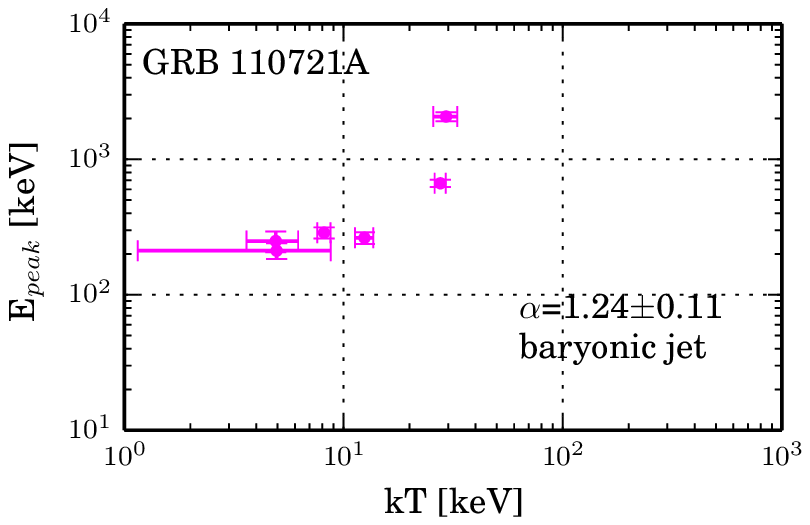}}\subfigure[]
{\includegraphics[scale=.7]{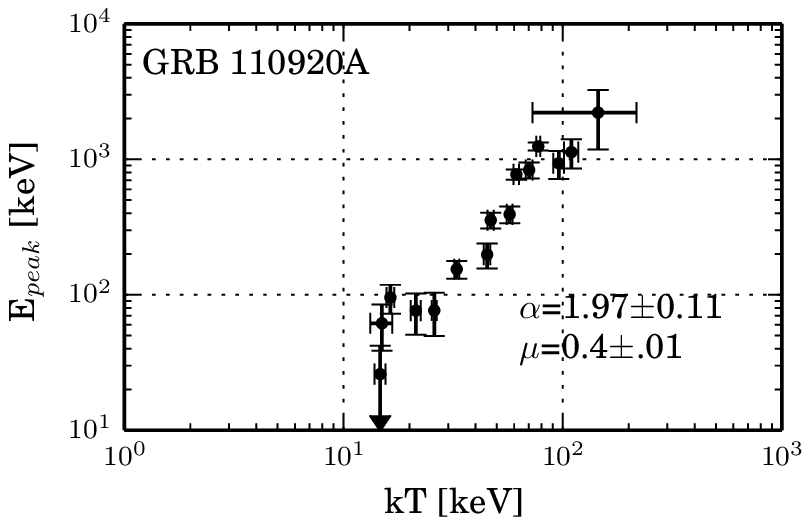}}\subfigure[]
{\includegraphics[scale=.7]{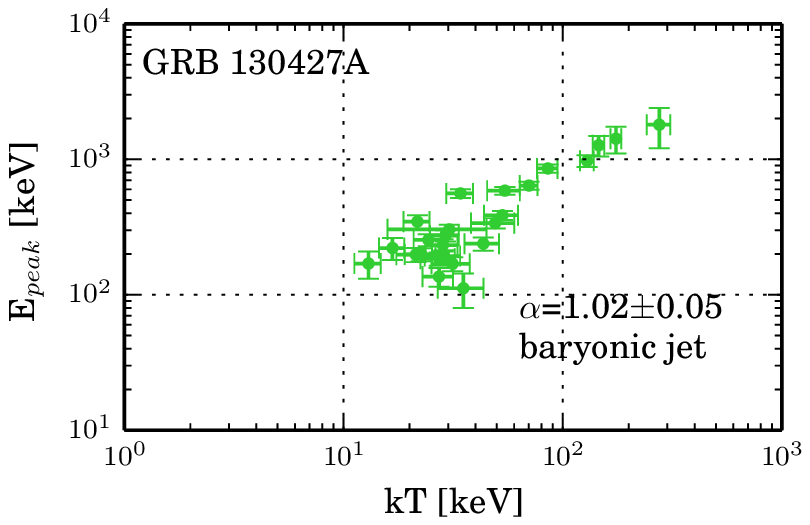}}
\caption{The individual observer frame correlations of each GRB all
  follow an observational relation of $E_{\rm p}\propto T^{\alpha}$
  where $\alpha$ ranges from $\sim1-2$. At low $E_{\rm p}$ and $kT$,
  the characteristic energies are less well-constrained in the weaker
  tails of some of the GRBs. The relatively fewer data points for GRB
  110721A (panel (d)) make the correlation difficult to measure.}
  \label{fig:cors}
\end{figure}

\begin{figure}[h]
  \centering
\includegraphics[scale=1.1]{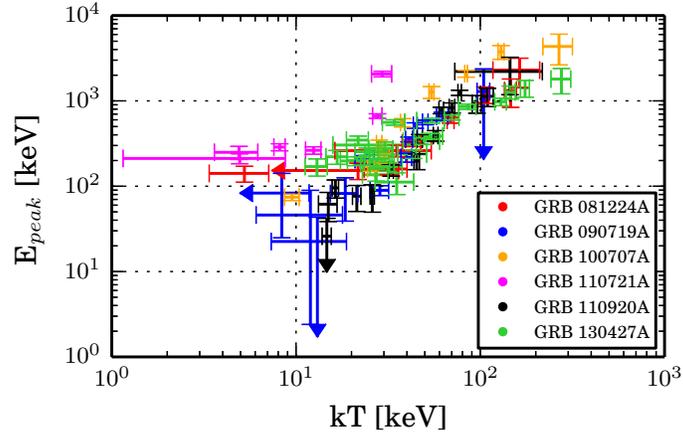}  
\caption{The time-resolved, observer-frame synchrotron and blackbody
  characteristic energies of the entire GRB sample. While the
  individual correlation index is not universal from burst to burst,
  it is clear that there is a strong correlation across the
  population. The Spearman rank correlation index, $\rho$, for the
  entire sample is 0.81 with a p-value of 4.35$\times10^{-20}$.}
\label{fig:allC}
\end{figure}

Figure \ref{fig:spec} shows an example of the spectral evolution of
the two separate components. A strong correlation is found between
$E_{\rm p}$ and $kT$, as illustrated in Figures \ref{fig:cors} \&
\ref{fig:allC}. A power law of the form $E_{\rm p}\propto T^{\alpha}$
was fit to the $E_{\rm p}$, $kT$ pairs of the individual GRBs yielding
values of $\alpha$ ranging from~$\sim$1 to 2 (see Table
\ref{tab:tab1}). The general temporal trend of both $E_{\rm p}$ and
$kT$ is an evolution from higher to lower energies. As can be seen
from Figure \ref{fig:spec}, the evolution of the flux of each
component is not necessarily tied to the change in the characteristic
energies. This is very evident during the rise phase of a pulse during
which the flux rises while $E_{\rm p}$ and $kT$ fall with
time. However, during the decay phase of the pulse, the flux decreases
along with the characteristic energies. Table \ref{tab:tab1} lists the
ratio of the blackbody flux to the total flux for each burst.



\section{Interpretation}

\edithere{To interpret these observations, we assume an emission
  process in which the thermal and non-thermal emission occur in
  close proximity to one another with the non-thermal synchrotron
  emission arising in an optically thin region above the photosphere
  of the jet. The range of the indices observed in the correlation
  suggests that the relation between the thermal and non-thermal
  emission varies from burst to burst. One way to achieve this is to
  assume that the composition of GRB outflows vary in their ratio of
  magnetic content from being magnetically to baryonically dominated.}
In this scenario, the jet dynamics are parameterized by the dependence
of the bulk Lorentz factor on the radius as $\Gamma\propto R^\mu$,
from its initial launching radius of $r_0$ until the jet reaches its
coasting Lorentz factor $\eta=L/\dot{M}c^2$ at the so-called
saturation radius $r_{\rm s}$, where L is the luminosity and $\dot{M}$
is the mass outflow rate. This will be approximately the jet's Lorentz
factor until it is decelerated upon collision with the surrounding
medium. For magnetically-dominated jets $\mu\approx 1/3$
\citep{Drenkhahn02,Drenkhahn:2002,Kirk:2003}, and in the baryonic case
$\mu\approx1$ \citep{Meszaros+93gasdyn}. Intermediate values
correspond to a mix of these components \citep{Veres+12fit}, and can
be further modified by factors such as the topology of the magnetic
field.

Under these assumptions, there are two regions of interest for which
we can define the radial evolution of the bulk Lorentz factor:
\begin{equation}
{\Gamma(r)}= \left\{
\begin{array}{lll}
 (r/r_0)^{\mu}	&	{\rm if} & r<r_{\rm sat}\\
 {\eta}	        &	{\rm if} & r_{\rm sat}<r
\end{array}
\right.
\label{eq:accel}
\end{equation}
Here, $r_{\rm sat}= r_0 \eta^{\frac{1}{\mu}}$ and is clearly larger
when the jet is magnetically dominated. The emission of the blackbody
is assumed to originate at the photospheric radius ($r_{\rm ph}$),
where the optical depth of the jet drops to unity. Following
\citet{Meszaros+93gasdyn}, the photospheric radius is
\begin{equation}
  \label{eq:pht1}
  \frac{r_{\rm ph}}{r_0}=\left(\frac{ L\sigma_{\rm T}}{8\pi m_{\rm p}
      c^3r_0}\right)\frac{1}{\eta \Gamma_{\rm ph}^2}
\end{equation}
where $\Gamma_{\rm ph}$ is the Lorentz factor of the outflow at
$r_{\rm ph}$. The value of $\Gamma_{\rm ph}$ depends on the magnetic
content of the outflow; therefore, $r_{\rm ph}$ can take on two
values,
\begin{equation}
\label{eq:pht2}
\frac{r_{\rm ph}}{r_0}=\eta_{\rm T}^{1/\mu}\left\{
\begin{array}{ll}
  (\eta_{\rm T}/\eta)^{1/(1+2\mu)}	& {\rm if~ } \eta>\eta_{\rm T} \\
  (\eta_{\rm T}/\eta)^3		& {\rm if~ } \eta<\eta_{\rm T}
\end{array}
\right. .
\end{equation}
The introduction of the critical Lorentz factor,
\begin{equation}
  \label{eq:critL}
  \eta_{\rm T}=\left(\frac{ L\sigma_{\rm T}}{8\pi m_{\rm p} c^3
      r_0}\right)^{{\mu}/{(1+3\mu)}}
\end{equation}
provides an important discriminator for the location of the $r_{\rm
  ph}$ relative to $r_{\rm s}$. Outflows with $\eta=\eta_{\rm T}$ have their
photospheres at the saturation radius. Typical observed Lorentz
factors of GRBs derived via different methods indicate values of a few
hundred \citep{Lithwick:2001,Peer:2007}. In a magnetically dominated
($\mu=1/3$) case, we have $\eta_{\rm T}\simeq 150~L_{53}^{1/6}
r_{0,7}^{-1/6}$. For physically relevant values of $L=10^{53} L_{53}$
erg s$^{-1}$ and $r_0=10^7r_{0,7}$ cm, $\eta_{\rm T}$ is low compared to
observed values for $\eta$. Therefore, the photosphere is in the
acceleration phase for a large segment of the parameter space. On the
other hand, in baryonic cases ($\mu=1$), $\eta_{\rm T}\simeq
1900~L_{53}^{1/4} r_{0,7}^{-1/4}$, which is several orders of
magnitude higher than observed Lorentz factors. Therefore, we assume
that magnetically dominated jets have their photospheres in the
acceleration phase and baryonically dominated jets have their
photospheres in the coasting phase. With this critical assumption, we
derive two cases for the behaviors of both $E_{\rm p}$ and $kT$.

Close above the photosphere, instabilities in the flow or magnetic
field line reconnection can lead to mildly relativistic shocks and
accelerate leptons, which in turn emit synchrotron radiation
\citep{Meszaros+11gevmag,McKinney+11switch}. \edithere{The synchrotron
  peak energy is dependent on the baryon number density
  $n'_b(r)=L/(4\pi r^2 m_p c^3 \Gamma(r) \eta)$ and the magnetic field
  $B'\propto {n'}_b^{1/2}$.  The peak synchrotron energy is: $E_{\rm
    p}=({3q_e B'_{\rm ph}}/ {4 \pi m_e c}) \gamma_{\rm e,ph}^2
  {\Gamma_{\rm ph}}$. From this expression we derive the following
  dependence on the input parameters:
\begin{eqnarray}
E_{\rm p} \propto \left\{
\begin{array}{ll}
  L^{\frac{3\mu-1}{4\mu+2}} \eta^{-\frac{3\mu-1}{4\mu+2}}
  r_0^{\frac{-5\mu}{4\mu+2}}		&	{\rm if~ }  \eta>\eta_{\rm T}\\
  L^{-1/2} \eta^{3} 
    		&	{\rm if~ }  \eta<\eta_{\rm T}.
\end{array}
\right.
\label{eq:ebreak}
\end{eqnarray}
} \edithere{The acceleration of the jet is assumed to be adiabatic for
  $r<r_{\rm sat}$, leading to a relation between the comoving
  temperature ($T^{\prime}$) and the comoving volume ($V^{\prime}$) of
  $T^{\prime}\propto V^{\prime -1/3}$. Since the expansion is along
  the radial direction of the jet, we have
  $V^{\prime}=d^{3}x^{\prime}=\Gamma d^{3}x \equiv 4\pi\Gamma r^2
  dr$. Using Equation \ref{eq:accel} for $r<r_{\rm sat}$, we can write
  $\Gamma\propto r^{\mu}$. Therefore, the comoving temperature of the
  sub-dominant thermal component depends on radius as ${T'}\propto
  r^{-\left(\frac{\mu+2}{3}\right)} $. Above the saturation radius,
  the standard evolution of the temperature is $T^{\prime}\propto
  r^{-2/3}$. At the launching radius ($r_0$) the temperature is
  $T_0=(L/4\pi r_0^2 a c )^{1/4}$. Therefore, the observed temperature
  for the two scenarios is:
\begin{eqnarray}
kT_{\rm obs}(r_{\rm ph}) \propto \left\{
\begin{array}{ll}
  L^{\frac{14\mu-5}{12(2\mu+1)}} \eta^{\frac{2-2\mu}{6\mu+3}}
  r_0^{-\frac{10\mu-1}{6(2\mu+1)}} 	& {\rm if~ } \eta>\eta_{\rm T} \\
  L^{-5/12} \eta^{8/3} r_0^{1/6}  	&	{\rm if~ } \eta<\eta_{\rm T} .
\end{array}
\right.
\label{eq:temp}
\end{eqnarray}
}




It is unclear whether the evolution of the photosphere's luminosity or
Lorentz factor, or some combination of both, drives the evolution of
$E_{\rm p}$ and $kT$.  One natural assumption is that the evolution of
the photosphere's luminosity results in the observed variations in
$E_{\rm p}$ and $T$ as a burst proceeds. Therefore, considering the
two types of jets; magnetically dominated ($r_{\rm ph}<r_{\rm s}$) and
kinetic dominated ($r_{\rm ph}>r_{\rm s}$) we have two possibilities
for the values of $\alpha$:

\begin{itemize}
\item \edithere{in the magnetic case}, considering the appropriate
  powers of $L$, we have $E_{\rm p} \propto
  T^{\frac{6(3\mu-1)}{14\mu-5}}$. The exponent is singular at
  $\mu\approx0.36$, but for values up to $\mu < 0.6$ (these are the
  values of $\mu$ for which the photosphere will occur in the
  acceleration phase) we are able to explain values of $\alpha$ from 2
  down to 1.4

\item \edithere{in the kinetic (baryonic)} case we have $E_{\rm p}
  \propto T^{1.2}$. This is observed in some GRBs.

\end{itemize}

\section{Discussion}

The analysis of GRBs in the framework of this model can indicate
whether the photosphere is in the acceleration or coasting phase,
which in turn can be translated to the composition of the jet.  We
find that for exponents close to 2 the jet dynamics are dominated by
the magnetic field while exponents close to 1 indicate baryonic
jets. In our sample of six GRBs observed with $\fermi$, the exponents
$\alpha$ of the relation between the characteristic energies of
non-thermal and thermal components (Table \ref{tab:tab1}) span the
range of possible values, showing that energy content of GRB jets
ranges from being dominated by the magnetic field to being contained
mostly in the kinetic energy of baryons in the jet. \edithere{A
  possible validation of this interpretation would be the future
  measurement of polarization in GRBs which will allow for the direct
  determination of the magnetization of GRB jets \citep[see for
  example][]{lundman:2013}.}

\edithere{ We note that the lack of a correlation between the ratio of
  the thermal flux to the total flux with the inferred magnetic
  content of the jet is puzzling (see Table \ref{tab:tab1}). Naively,
  it is expected that a photosphere occurring deep in the acceleration
  phase of the outflow will have its thermal emission be much brighter
  than the non-thermal emission. A possible explanation for the
  weakness of the observed thermal component has been addressed by
  several authors \citep{zhang:2011,daigne:2013}. These works consider
  the effect of the magnetization parameter ($\sigma=
  \frac{B^2}{4\pi\Gamma\rho c^2}$) on the intensity of the thermal
  component where $\rho$ is the matter density of the outflow. For
  $\sigma\gg1$, most of the jet internal energy remains in the
  advected magnetic field, reducing the intensity of the observed
  thermal component from the photosphere. Another possibility for
  explaining the lack of correlation of the thermal flux ratios to the
  different jet modes is to consider that if the non-thermal flux is
  due to synchrotron following reconnection events above the
  photosphere, the amount of reconnection may not be simply given by
  the amount of magnetic energy and by the radius, but may depend also
  on the degree of tangledness of the field at that radius. For
  reconnection one needs field lines of opposite polarity near each
  other, and if the degree of randomness is stochastic (as it probably
  is), this could introduce a randomness in the amount of non-thermal
  electrons accelerated as well as the synchrotron flux
  produced. However, time-dependent simulations of magnetically
  dominated outflows in GRBs are not advanced enough to accurately
  test these assumptions and therefore the reduced intensity of the
  thermal component is still open to interpretation. }

\acknowledgements The $\fermi$ GBM collaboration acknowledges support
for GBM development, operations and data analysis from NASA in the US
and BMWi/DLR in Germany. We also thank the anonymous referee for very
useful comments that aided in refining this work.


\clearpage

\begin{table}
\centering
\begin{tabular}{c || c | c | c | c }
 
    GRB Name   &           $\alpha$  & Jet Type & $\mu$  & $F_{BB}/F_{tot}$ \\\hline\hline
  GRB 081224A  &	$1.01   \pm	0.14	 $ & baryonic   &   $-$          &    0.29\\
  GRB 090719A   &	$2.33   \pm	0.27	 $ & magnetic   & 0.39$\pm$0.01  &    0.27\\
  GRB 100707A  &	$1.77	\pm	0.07	 $ & magnetic   & 0.42$\pm$0.01  &    0.33\\
  GRB 110721A&	$1.24	\pm	0.11	 $ & baryonic   &     $-$        &    0.01\\
  GRB 110920A                    &	$1.97	\pm	0.11	 $ & magnetic   & 0.4$\pm$0.01   &    0.39\\
  GRB 130427A &	$1.02	\pm	0.05	 $ &  baryonic  &     $-$        &    0.22\\

\end{tabular}
\caption{Indices ($\alpha$) and derived Lorentz factor radial indices ($\mu$) from fitting power laws to the $E_{\rm p}$, $kT$ pairs for each GRB. The inferred jet type and blackbody (F$_{BB}$) to total flux (F$_{tot}$) 
  ratios are also given. The values of $\alpha$ for the GRBs vary but are within the 
  constraints of the model derived via our interpretation. Values of $\mu$ are listed only for those GRBs that are inferred to be magnetically dominated. No significant correlation 
  between $\alpha$ and the flux ratios was found in the data. }
\label{tab:tab1}
\end{table}

\end{document}